\documentclass[11pt]{article}
\newcommand{\Comment}[1]{{}}
\usepackage{amsfonts,amsmath, amsmath,amssymb,slashed}
\usepackage[textwidth = 430 pt, textheight = 630 pt]{geometry}

\Comment{\usepackage{color}
\definecolor{MyDarkBlue}{rgb}{0.15,0.15,0.45}
\usepackage[linktocpage=true]{hyperref}
\hypersetup{
colorlinks=true,
citecolor=MyDarkBlue,
linkcolor=MyDarkBlue,
urlcolor=MyDarkBlue,
pdfauthor={},
pdftitle={},
pdfsubject={hep-th}
}

\usepackage[numbers,sort&compress]{natbib}
\usepackage{hypernat}}
\usepackage{graphicx}
\usepackage{cite,color}

\def\Tr{{\rm Tr\, }}
\def\e{{\rm e}}
\def\ep{\epsilon}
\def\a{\alpha}

\def\l{\lambda}

\def\d{\partial}
\def\dag{\dagger}

\def\CP{\mathbb{CP}}

\newcommand{\be}{\begin{equation}}
\newcommand{\bea}{\begin{eqnarray}}

\newcommand{\ee}{\end{equation}}
\newcommand{\eea}{\end{eqnarray}}

\newcommand{\nn}{\nonumber}

\begin{document}

\renewcommand{\thefootnote}{\fnsymbol{footnote}}

\makeatletter
\@addtoreset{equation}{section}
\makeatother
\renewcommand{\theequation}{\thesection.\arabic{equation}}

\rightline{}
\rightline{}


\vspace{10pt}


\begin{center}
{\LARGE \bf{\sc Open strings on D-branes from ABJM}}
\end{center} 
 \vspace{1truecm}
\thispagestyle{empty} \centerline{
{\large \bf {\sc Carlos Cardona${}^{a,}$}}\footnote{E-mail address: \Comment{\href{mailto:cargicar@ift.unesp.br}}{\tt cargicar@ift.unesp.br}}and
{\large \bf {\sc Horatiu Nastase${}^{a,}$}}\footnote{E-mail address: \Comment{\href{mailto:nastase@ift.unesp.br}}{\tt nastase@ift.unesp.br}} 
                                                           }

\vspace{.5cm}

\centerline{{\it ${}^a$ 
Instituto de F\'{i}sica Te\'{o}rica, UNESP-Universidade Estadual Paulista}} \centerline{{\it 
R. Dr. Bento T. Ferraz 271, Bl. II, Sao Paulo 01140-070, SP, Brazil}}

\vspace{1truecm}

\thispagestyle{empty}

\centerline{\sc Abstract}

\vspace{.4truecm}

\begin{center}
\begin{minipage}[c]{380pt}
{We study open strings on giant gravitons (D-branes on cycles) in the ABJM/$AdS_4\times \mathbb{CP}^3$ correspondence. We find that their energy 
spectrum has the same form as the one for closed strings, with a nontrivial function of the coupling, avoiding BMN scaling. A similar, Cuntz oscillator,
Hamiltonian description for the string (operators at strong coupling) to the $AdS_5$ case is valid also in this case.
}
\end{minipage}
\end{center}

\vspace{.5cm}

\setcounter{page}{0}
\setcounter{tocdepth}{2}


\renewcommand{\thefootnote}{\arabic{footnote}}
\setcounter{footnote}{0}

\linespread{1.1}
\parskip 4pt


\newcommand{\bean}{\begin{eqnarray*}}
\newcommand{\eean}{\end{eqnarray*}}


\section{Introduction}

The ABJM/$AdS_4\times \mathbb{CP}^3$ correspondence \cite{Aharony:2008ug} has received a lot of interest. Many things work as in the original case of 
${\cal N}=4$ SYM vs. $AdS_5\times S^5$, but there are important differences. In particular, there is a spin chain description and a corresponding Bethe 
ansatz for ABJM operators \cite{Minahan:2008hf,Gaiotto:2008cg,Gromov:2008qe}, but the energy of a magnon does not have the formula compatible with 
BMN scaling of operators of large charge ($\lambda/J^2=g^2N/J^2$ in ${\cal N}=4$ SYM), 
and instead one has a nontrivial function $h(\lambda)$ of the 't Hooft coupling $\lambda=N/k$, giving \cite{Gaiotto:2008cg,Nishioka:2008gz,Grignani:2008is}
\be
\epsilon(p)=\frac{1}{2}\sqrt{1+16h^2(\lambda)\sin^2\frac{p}{2}}\;,\label{magnondisp}
\ee
where $p$ is the magnon momentum, which is $=2\pi k/J\ll 1 $ in the BMN limit, obtaining $\epsilon(p)=1/2\sqrt{1+16\pi^2h^2(\lambda)k^2/J^2}$. The function 
$h(\lambda)$ is nontrivial, and whereas at strong coupling we have $h(\lambda\gg 1)\simeq \sqrt{\lambda/2}+a_1+...$ as in the BMN case, at weak coupling we have 
$h(\lambda)=\lambda(1+c_1\lambda^2+c_2\lambda^4+...)$. The subleading term at strong coupling, $a_1$, was the subject of some debate, since it depends on the 
method of regularization for quantum worldsheet corrections (see e.g. \cite{McLoughlin:2008he}). In \cite{LopezArcos:2012gb} it was argued that imposing a 
physical principle can help define a regularization and a unique value for $a_1$.

It is then important to consider other ABJM excitations than the closed strings corresponding to the magnons above. In theories with objects in the fundamental 
of the gauge group we can define open strings, as first introduced in \cite{Berenstein:2002zw}. But one can define also other kinds of open strings, in 
particular ones that extend between D-branes. In the context of ${\cal N}=4$ SYM, these have been studied in \cite{Berenstein:2005vf,Berenstein:2005fa,Berenstein:2006qk}. 
D-branes wrapping some cycles in the gravity dual, with some angular momentum, are usually giant gravitons, i.e.
states with the momentum of a graviton, but extended in space. 

In this paper we therefore study open strings with large angular momentum attached to maximal giant gravitons 
in $AdS_4\times \mathbb{CP}^3$, and their dual description in ABJM, as 
gauge invariant operators with large R-charge. These can be thought of as excitations of the giant gravitons.
Giant gravitons in $AdS_4\times \mathbb{CP}^3$ have been studied among others in 
\cite{Giovannoni:2011pn,Lozano:2013ota} and references therein. Within M theory, large R-charge states and their gravity dual have 
been studied in \cite{Kovacs:2013una}.

We will study the anomalous dimensions of operators dual to these open strings on giant gravitons, and we will 
find, perhaps not surprisingly, that the same formula valid for the magnon dispersion relation (\ref{magnondisp}) is valid now, at 
least for the leading terms at weak coupling $\lambda\ll 1$ and strong coupling $\lambda \gg 1$.

In \cite{Murugan:2013jm, Mohammed:2012rd, Mohammed:2012gi}, an abelian reduction of the massive ABJM model down to a Landau-Ginzburg system relevant for 
condensed matter physics was described, but the reduction of the gravitational dual was not well understood. It seems however that if we consider operators 
of large charge, we should obtain a pp wave in the gravity dual, and possible relevant states include giant gravitons on the pp wave and their excitations. 
Therefore it is of great interest to understand the physics of giant gravitons on pp waves and their excitations.

The paper is organized as follows. Section 2 reviews the giant gravitons we are considering from ABJM, D4-branes wrapping a $\mathbb{CP}^2\subset\mathbb{CP}^3$
and the eleventh dimension. In section 3 we describe the Penrose limit of the $AdS_4\times \CP^3$ background relevant for our calculation, and the quantization 
of open strings on D-branes in the resulting pp wave background. In section 4 we describe the construction of operators corresponding to giant gravitons with 
open strings on them, and with excited open string states. In section 5 we calculate the first, 2-loop correction to the anomalous dimension of open string 
operators with one impurity, and compare with the string theory side. In section 6 we describe the Hamiltonian analysis using Cuntz oscillators that gives
the complete perturbative result (resumming the 2-loop result and being valid at arbitrary magnon momentum), and in section 7 we conclude. 
In Appendix A we show the details of how the analysis of the Hamiltonian for BMN operators works in ${\cal N}=4$ SYM using Cuntz oscillators, which is paralleled
in our case. In Appendix B we give the details of the 2-loop calculation for the anomalous dimension.

\section{Review of giant gravitons in $AdS_4\times \CP^3$.}

In this section we summarize a particular embedding of a giant graviton in $AdS_4\times \CP^3$. We consider a D4-brane 
wrapping a submanifold of ${\CP}^3$, stabilized by the presence of the four-form flux, balancing their tension.
It is well-known that a large class of M5-branes, known as sphere giant gravitons, can be embedded into the background $AdS_ 4\times S^7$ of eleven dimensional supergravity.
Several configurations of this kind were constructed in \cite{Mikhailov:2000ya} for different backgrounds and with different amount of preserved supersymmetries.

In there, by embedding an $S^7$ in $\mathbb{C}^4$ and considering intersections with holomorphic surfaces in
$\mathbb{C}^4$, it was possible to obtain configurations preserving $1/2,\,1/4,\,$ and $1/8$ supersymmetries. More explicitly, the worldsurface of the giant graviton at time $t$ is described by the following constraints in $\mathbb{C}^4$,
\bea
\sum_{M=1}^4 |Z_M|^2=1\,,&&\nn\\
F(\e^{-i\,t/R}Z_M)=0&&M=1..4\,.\label{holofun}
\eea
Spherical giant gravitons of the form $F(Z_M)=0$ preserve 1/2 of the supersymmetry (a particular solution of this type has been considered in \cite{Herrero:2011bk}), those of the form $F(Z_M,Z_N) = 0$ preserve
1/4 and those given by $F(Z_M, Z_N, Z_P) = 0$ and $F(Z_M, Z_N, Z_P, Z_Q) = 0$  preserve 1/8 (solutions of this type have been considered in \cite{Lozano:2013ota}.
See \cite{Abbott:2013ija} for a classification of the topology of giant gravitons).
We are particularly interested in those given by holomorphic curves of the type  $F(Z_N,Z_M)=0$,
which after the dimensional reduction to $IIA$ string theory wrap a subspace of ${\CP}^3$.

\subsection{D4-brane giant graviton on $\CP^3$}

We consider an $M5-$brane wrapping an $S^5\subset S^7$ in the $AdS_4\times S^7/\mathbb{Z}_k$ background of eleven dimensional supergravity. 

The background is given by
\be
ds^2= R^2\,(ds^2_{AdS_4}+ds^2_{S^7/\mathbb{ Z}_k})\,,
\ee
with the radius of the sphere $R=(32\pi^2Nk)^ {1/6}$
in string units. 
Since $S^7$ is an $S^1$ fibered over $\CP^3$, we can write the $S^7/\mathbb{Z}_k$ metric as
\begin{equation}
ds^2_{S^7/{\mathbb{Z}_k}}=\left(\frac{1}{k}d\tau+\mathcal{A}\right)^2+ds^2_{\CP^3}\;,\label{s7cp3}
\end{equation}
where $\tau\in[0,2\pi]$. 

To compute the explicit form for this metric decomposition, we start with the metric of $\mathbb{C}^4$ \cite{Pope:1980ub, Pope:1984bd},
\be\label{C4}
ds^2= \sum_{M=1}^4|dZ_M|^2\;,
\ee
and restrict to $S^7$ by the constraint
\be\label{S7}
\sum_{M=1}^4|Z_M|^2=1.
\ee 

To restrict further to $S^7/{\mathbb Z}_k$, or $\mathbb{CP}^3$ in the limit $k\rightarrow \infty$ relevant for AdS/CFT, we
impose the equivalence $Z_M\sim Z_M\e^{i\alpha}$, where $\a=2\pi/k$ for $S^7/{\mathbb Z}_k$ and $\a$ is arbitrary for $\mathbb{CP}^3$.
 
In other words, $\CP^3$ is the the space of orbits under the action of $U(1)$ on the homogenous coordinates $Z_i$. We can then forget the 
constraint (\ref{S7}) which becomes irrelevant as we rescale all the $Z_i$ by an arbitrary quantity, and think of the 
$Z_i$ as homogenous coordinates for the $\CP^3$.
We can define inhomogenous (affine) coordinates on $\CP^3$ that get rid of the $U(1)$ ($S^1$) fiber coordinate $\tau$ from the point of view of $S^7$,
\be\label{AffineCp3}
\zeta_l=\frac{Z_l}{Z_4},\,Z_4=|Z_4|\e^{i\tau} ,\quad l=1,2,3\;,
\ee
in terms of which the metric of $S^7$ is given by
\be
ds^2_{S^7}=(d\tau+\mathcal{A})^2+\frac{\sum_{l}|d\zeta_l|^2}{(1+\sum_l|\zeta_l|^2)}-\frac{\sum_{l,k}\zeta_l\bar{\zeta}_k d\zeta_l d\bar{\zeta}_k}{(1+\sum_l|\zeta_l|^2)^2}\,,
\ee
which is the sought-for explicit form of (\ref{s7cp3}).
Here $\mathcal{A}=\frac{i\sum_{l}\bar{\zeta}_l d\zeta_l-c.c}{2(1+\sum_l|\zeta_l|^2)}$, and the metric is known as the Fubini-Study metric. 
By dropping the first term in the above equation, we get the metric of $\CP^3$. 

It is also useful to introduce a real six-dimensional metric on $\CP^3$ by defining 6 angles as
\bea\label{affinecp3}
\zeta_1 &=&\tan\mu\sin\alpha\sin(\theta/2)\,\e^{i(\psi-\phi)/2}\,\e^{i\chi/2} ,\nn\\ \zeta_2 &=&\tan\mu\cos\alpha\,\e^{i\chi/2},\nn\\ \zeta_3 &=&\tan\mu\sin\alpha\cos(\theta/2)\,\e^{i(\psi+\phi)/2}\,\e^{i\chi/2}\,,
\eea
in terms of which the metric on $\CP^3$ is
\begin{eqnarray}
\label{theCP^3}
ds_{\CP^3}^2&=&d\mu^2+\sin^2\mu\,\Big[ d\alpha^2+\frac{1}{4}\sin^2\alpha\, \big(\sigma_1^2+\sigma_2^2+\cos^2\alpha\,\sigma_3^2\big)+\frac{1}{4}\cos^2\mu\, \big(d\chi+\sin^2\alpha\, \sigma_3\big)^2\Big].\cr
&&
\end{eqnarray}
Here $\sigma_{1,2,3}$ are left-invariant 1-forms on an $S^3$, given explicitly by
\begin{eqnarray}
\sigma_1&=&\cos\psi\, d\theta+\sin\psi\sin\theta\,d\phi\,,\nn\\
\sigma_2&=&\sin\psi\, d\theta-\cos\psi\sin\theta\,d\phi\,,\nn\\
\sigma_3&=&d\psi+\cos\theta\,d\phi.
\end{eqnarray}

The range of the 6 angles is
\begin{equation}
0\le \mu,\,\alpha\le \frac{\pi}{2}\, ,\quad 0\le \theta \le \pi\, ,\quad 0\le \phi\le 2\pi\, ,\quad 0\le \psi,\, \chi\le 4\pi\ ,
\end{equation}
and the 1-form defining the embedding in M-theory is
\begin{equation}
\mathcal{A}=\frac{1}{2}\sin^2\mu\, \Big(d\chi+\sin^2\alpha\,\big(d\psi-\cos\theta\, d\phi)\Big)\, .\label{theA}
\end{equation} 

We could similarly write an $S^5$ as an $S^1$ bundle over the $\CP^2$ by 
\begin{equation}
\label{elCP^2}
ds^2_{S^5}=(d\chi'+A)^2+ds^2_{\CP^2}. 
\end{equation}
The $\CP^2$ is written in terms of the $\CP^3$ angles as 
\be
ds^2_{\CP^2}= d\alpha^2+\frac{1}{4}\sin^2\alpha\, \big(\sigma_1^2+\sigma_2^2+\cos^2\alpha\,\sigma_3^2\big).\label{cp2}
\ee
From (\ref{theCP^3}) and (\ref{cp2}) we see the embedding of $\CP^2\subset\CP^3$. 
Specifically, we have 
\be
ds^2_{\CP^3}=d\mu^2+\sin^2\mu\left[\cos^2\mu(d\chi'+A)^2+ds^2_{\CP^2}\right].\label{cp2cp3}
\ee
(\ref{elCP^2}) means that the embedding of $\CP^2\subset S^5$ is also a $S^1$ fibration, like the embedding of $\CP^3\subset S^7$ in (\ref{s7cp3}).

We would like to consider the brane whose spatial components are wrapping the subspace 

\be
\label{d4} ds^2_{\CP^3}\big(\mu=\frac{\pi}{4},\alpha=0\big) \,.
\ee

We are unsure whether this brane can be described using a holomorphic function (which would immediately imply supersymmetry) as in 
the discussion following (\ref{holofun}).\footnote{We would like to thank Andrea Prinsloo for pointing out to us that a proposal for a function $F$ 
that appeared in the first version of the paper was incorrect.}

\section{PP-wave limit and open strings in $AdS_4\times {\mathbb CP}^3$.}

We consider composite operators in ABJM carrying large $R$-charge $J$. As we have learned from the BMN case\cite{Berenstein:2002jq}, the states with
$J^2\sim \lambda$ are well described in the gravitational side by a Penrose limit of the gravitational background.
In this section we shall consider the Penrose limit of type IIA background $AdS_4\times \CP^3$.

\subsection{Penrose limit along the giant.}

The metric of $AdS_4$ in global coordinates is
\be 
ds^2_{AdS_4}=R^2\left(-{\rm cosh}^2\rho\,dt^2+d\rho^2+{\rm sinh}^2\rho\,d\Omega_2^2\right)\,,
\ee
and the metric on $\CP^3$ was given in (\ref{theCP^3}).

We would like to take the Penrose limit defined by focusing on the geodesic propagating at the speed of light along $\chi$, 
with $\mu=\pi/4\,,\alpha=0$ and $\rho=0$, which corresponds to the position of the maximal giant graviton. 
The giant graviton wraps $(\psi,\theta,\phi,\chi)$, so the Penrose limit is {\em along } the giant, since we want to describe open string states 
propagating in $\chi$, not giant graviton states. The Penrose limit is then defined by the transformations
\be 
\rho=\frac{\tilde{\rho}}{R},\quad\mu=\frac{\pi}{4}+\frac{u}{R},\quad\alpha=\frac{r}{R},\quad x^+=\frac{t+\chi/2}{2},\quad x^-=R^2\frac{t-\chi/2}{2}\,,
\ee
followed by taking $R\to \infty$. The metric then reduces to \cite{Gaiotto:2008cg}
\bea
ds^2 &=& -4dx^+dx^- + du^2 + d\tilde\rho^2 + \tilde\rho^2 d\Omega_2^2 + dr^2 + {r^2\over 4}\sum_{i=1}^3\sigma_i^2
-(u^2+\tilde\rho^2)(dx^+)^2 +{1\over 2}r^2\sigma_3 dx^+ \cr
&=& -4dx^+dx^- + du^2 + \sum_{i=1}^3 dy_i^2  + \sum_{a=1}^2 dz_a d\bar z_a
-\left(u^2+\sum_{i=1}^3 y_i^2\right)(dx^+)^2\cr
&& -  {i\over 2}\sum_{a=1}^2 (\bar z_a dz_a - z_a d\bar z_a) dx^+\;,
\eea
where $y_i$, $i=1,2,3$ are cartesian coordinates 
for the spherical coordinates $\tilde \rho, \Omega_2$, and
$z_1$, $z_2$ are complex coordinates on $\mathbb{C}^2$ with spherical coordinates $(r,\theta,\phi,\psi)$.
After a further coordinate change
\begin{equation}
z_a = e^{-{i}x^+/2} w_a,~~~~\bar z_a = e^{{i}x^+/2} \bar w_a,\label{x+rot}
\end{equation}
the metric takes the standard pp-wave form (with an extra term),
\bea\label{ppmetric}
ds^2 &=& -4dx^+dx^- + du^2 + \sum_{i=1}^3 dy_i^2  + \sum_{a=1}^2 dw_a d\bar w_a
-\left(u^2+\sum_{i=1}^3 y_i^2 + {1\over 4}\sum_{a=1}^2 |w_a|^2\right)(dx^+)^2\cr
&&-\frac{i}{2}\sum_{a=1,2}\left(w_ad\bar w_a-\bar w_a d w_a\right)dx^+.
\eea
The extra term can be absorbed by a transformation in $x^-$ such as,
\be
dx^ -\rightarrow dx^ --\frac{i}{8}\sum_{a=1,2}\left(w_ad\bar w_a-\bar w_a d w_a\right)\,.
\ee
The fluxes reduces to
\begin{equation}
\begin{aligned}
& F_2 = - dx^+\wedge du, \\
& F_4 = -3 dx^+ \wedge dy_1 \wedge dy_2 \wedge dy_3.
\end{aligned}
\end{equation}

The brane in this background is wrapping the light-cone directions plus an $S^3$ embedded in the four-dimensional space spanned by ($w_1,w_2$).
Note that the Penrose limit breaks the isometry group $SO(2,4)\times (SU(4)/Z(SU(4))$ of $AdS_4\times \mathbb{CP}^3$ 
down to $U(1)_{\pm}\times U(1)_u\times SO(3)_r\times SO(3)\sim U(1)_{\pm}\times U(1)_u\times SU(2)_r\times SU(2)_L$. Here
$U(1)_{\pm}$ corresponds to $x^\pm$ boosts (coming from boosts along the compact coordinate $\chi$), $U(1)_u$ to translations along the (compact) coordinate
$u$, $SU(2)_r$ rotates $(y_1,y_2,y_3)$, whereas $SU(2)_L$ is the rotation group acting on the complex coordinates $(w_1,w_2)$. Note that if $(w_1,w_2)$ 
were written as 4 real coordinates, there would be an $SO(4)=SU(2)_L\times SU(2)_R$ action on them, but now $SU(2)_R$ is broken to its Cartan generator
(with rotations $e^{+i\a}$ and $e^{i\a}$ on the diagonal), identified with $U(1)_\pm$ because of (\ref{x+rot}).

In the pp-wave background, i.e. after the Penrose limit, the open string attached to the D4-brane is moving with $U(1)_\pm$ angular momentum given by $J_{\chi}=-i\partial_{\chi}$, 
and in global $AdS_4$ coordinates its energy is given by $E=i\partial_t$. In terms of the ABJM theory, these should corresponds to the conformal weight $\Delta$ and R-charge under a particular $U(1)$, for a state of the field theory on $S^2\times \mathbb{R}$. 
The pp wave light-cone momentum $2p^+$ and energy $2p^-$ of the open string are related to the $AdS_4$ quantities as follows.
\be\label{rch}
2p^+=\frac{i(\partial_t-\partial_{\chi})}{\tilde{R}^2}=\frac{\Delta+J_{\chi}}{{\tilde{R}^2}},\quad 2p^-=i(\partial_t+\partial_{\chi})=\Delta-J_{\chi}\,.
\ee
Here $\tilde{R}^2=2^{5/2}\pi\sqrt{\lambda}$ is the radius in string units in terms of the ABJM quantities. 
A state spinning with finite $p^+$ should correspond
in the field theory to a state with R-charge of order $J_{\chi}\sim \tilde R^2\sim \sqrt{\lambda}$. Since $k$ is an integer, the largest coupling corresponds to $k=1$, which suggests that the maximal charge for operators should be of order $J_{\chi}\sim \sqrt{N}$. We will come back to this issue after we define states 
in the field theory dual to the pp wave.

\subsection{Quantum open string on the pp wave}

The quantization of open strings in pp-waves backgrounds and its relation to CFT operators has been described in \cite{Berenstein:2002zw}.

In the context of D$-p$-branes it was considered in \cite{Dabholkar:2002zc}. 
In the light cone gauge $x^+=\tau$, $\Gamma^+\Theta=0$, the Green-Schwarz action
for the type IIA string is, following the conventions of \cite{Michelson:2002ps}), (see also \cite{Sugiyama:2002tf},\cite{Hyun:2002wu})
\begin{equation}
\begin{aligned}
S &= {1\over 4\pi\alpha'}\int dt\int_0^{\pi\alpha' p^+} d\sigma \left\{\sum_{A=1}^8 \left[ (\dot X^A)^2 -({X^A}')^2 \right]
-\sum_{M=1}^4 (X^M)^2 - {1\over 4}\sum_{N=5}^8 (X_N)^2\right. \\
&\left.~~~~~~
-{i\over 2} \bar \Theta\Gamma^-\left[\partial_\tau+\Gamma^{11}\partial_\sigma
-{1\over 4} \Gamma^1\Gamma^{11} -{3\over 4}\Gamma^{234} \right]\Theta
\right\}\,.\label{stringppaction}
\end{aligned}
\end{equation}
Here we have denoted $(u,y_i)=(X^M),\,\,M=1\cdots 4$ and $(w_a,\bar{w}_a)=(X^N),\,\,N=5\cdots 8$.

The open string we are interested in ends on a D4-brane wrapping the space spanned by $(x^{\pm},X_N)$. It follows that we should impose  
Neumann boundary conditions on the directions  $X^N,\, N=5,6,7,8$ and Dirichlet boundary conditions for the remaining coordinates,
\be 
\d_{\sigma}{X}^{N}=0\,\quad\text{for}\,\,j=5,6,7,8.\,\quad\d_{\tau} X^{M}=0,\,\text{for}\,\,M=1,2,3,4\,.
\ee
The bosonic excitations of the type IIA string in this pp-wave background have light cone
spectrum
\begin{equation}\label{bosexc}
H = \sum_{M=1}^4\sum_{n=-\infty}^\infty N_n^{(M)}\sqrt{1+{n^2\over(\alpha'p^+)^2}}+\sum_{N=5}^8 
\sum_{n=-\infty}^\infty N_n^{(N)}\sqrt{{1\over 4}+{n^2\over (\alpha'p^+)^2}}.
\end{equation}
In terms of the ABJM gauge theory variables, we have $R^2/\alpha' = 2^{5/2}\pi \sqrt{\lambda}$ and
$p^+=J/R^2$.
 Thus we have 4 excitations of frequency $1/2$ at $n=0$, corresponding to $(w_a,\bar w_a)$, and 4 excitations of frequency $1$ at $n=0$, corresponding to 
$(y_i,u)$. In this paper we will focus on the 4 excitations of frequency 1/2 at $n=0$. 

The Green-Schwarz action has bosonic symmetry group \cite{Gaiotto:2008cg} $SU(2)_{i}\times U(1)\times SO(4)$, where $SO(4)$ corresponds to rotations 
along the directions of the worldvolume of the D4-brane $X^N,\,N=5,6,7,8$, and $SU(2)_{r}$ corresponds to rotations in the 3 directions transverse 
to the brane  $i=1,2,3$ (Note that $X^{M}$ splits as $(u,y_i),\quad i=1,2,3$, and the action (\ref{stringppaction}) has only $SO(3)=SU(2)$ symmetry, 
not $SO(4)$, because of the fermionic part).
 
The vacuum of the string should be chosen such that it is invariant under rotations transverse to the giant and has a given charge $q'$ under $SU(2)_{i}$.

\section{Open strings from operators in ABJM}

Type IIA strings on $AdS_4\times \CP^3$ has been argued in \cite{Aharony:2008ug} to be dual to ${\cal N}=6$ Chern-Simons matter theory in three dimension 
with level $(k,-k)$ and gauge group $U(N)\times U(N)$. The theory becomes weakly coupled when the level $k$ is large, hence in the large $N$ limit the 
coupling analogous to 't Hooft coupling is given by $\l=N/k$, which is kept finite. The gauge fields are coupled to four chiral superfields 
in the bifundamental representation of the gauge group  $U(N)\times U(N)$, and in the fundamental representation of the $SU(4)$ R-symmetry. 
We denote the complex scalars in these 4 chiral multiplets as $(A_1,A_2,\bar{B}_1,\bar{B}_2)$. Here $A_1,A_2$ are in the $(N,\bar N)$ representation of $U(N)\times U(N)$, 
whereas $B_1,B_2$ are in the conjugate, $(\bar N, N)$. Under the $SU(4)_R$ R-symmetry group ($A_1,A_2, \bar{B}_1,\bar{B}_2$) transform in the $\mathbf{4}$
representation.
There is also a $U(1)_R$ under which all of $(A_1,A_2,\bar B_1, \bar B_2)$ have charge $+1$.

Differentiating between $A_i$ and $B_i$, for instance by adding a mass deformation the ABJM Lagrangean breaks $SU(4)_R$ to $SU(2)_A\times SU(2)_B\times U(1)$, 
under which $A_i$ transform as $(2,1,+1)$, i.e. a doublet under $SU(2)_A$, singlet under $SU(2)_B$ and charge $\tilde J(A_i)=+1$ under $U(1)$; and 
$B_i$ transform as $(1,2,-1)$, i.e. singlet under $SU(2)_A$, doublet under $SU(2)_B$ and charge $\tilde J(B_i)=-1$ under $U(1)$.

Note however that there is another possible breaking of $SU(4)_R$ which will turn out to be relevant for us, namely to $SU(2)_G\times SU(2)_{G'}\times U(1)'$.
Under this breaking, $(A_1,\bar B_1)$ transforms as a doublet of $SU(2)_G$ and a singlet of $SU(2)_{G'}$, and have $U(1)'$ charge $J'(A_1,B_1)=+1$; and 
$(A_2,\bar B_2)$ transform as a doublet of $SU(2)_{G'}$ and a singlet of $SU(2)_{G}$ and have $U(1)'$ charge $J'(A_2,\bar A_2)=-1$.

Analogously to the ${\cal N}=4$ SYM case, a giant graviton brane wrapping some cycle in the background should be identified with a semi-determinant of scalars 
\cite{Balasubramanian:2001nh}, but since the scalars $A_a,B_a$ carry indices in different $U(N)$ on the left and the right, unlike in ${\cal N}=4$ SYM, 
we should build composite fields which carry indices in the same $U(N)$ on both sides, i.e. in the adjoint of one of the $U(N)$'s. 
If such giant gravitons wrap the compact sector $\CP^3$ (as opposed to the $AdS_4$ piece, which also has its giant gravitons), their angular momentum 
is bounded from above due to the finite radius of $\CP^3$, 
and the maximal giant should be described in the field theory by a full determinant. We are particularly interested in the 
maximal giant graviton wrapping a subspace ${\cal M}_4\subset \CP^3$.

We would like to obtain the open string spectrum (\ref{bosexc}) in the field theory from  a string of composite operators in the adjoint of $U(N)$. Analogously to the closed string case \cite{Gaiotto:2008cg}, let us choose the vacuum of the string as,
\be
W^a_b=[(A_2B_2)^J]^a_b \,.
\ee
It should correspond to a zero energy configuration above the energy of the D4-brane $2p^-=H=0$ , which means in the field theory we need
\be\label{jchi} 
\big(\Delta-J_{(A_2\bar{B}_2)}\big)\big[(A_2B_2)\big]=0\,,
\ee
$\Delta$ being the conformal dimension, which classically is 1/2 for both $A_a$ and $B_a$. 
In order to have $J=1$ for $A_2B_2$, we need $J$ to be $J_{(A_2\bar{B}_2)}$, i.e. 
the Cartan generator of the $SU(2)_{(A_2\bar{B}_2)}=SU(2)_{G'}$ which rotates $(A_2\bar{B}_2)$ as a doublet, with (normalized) charge $+1/2$ for $A_2$ and 
$-1/2$ for $\bar B_2$, so $+1/2$ for $B_2$. 

On the other hand, the vacuum operator $W^a_b$ is invariant under the action of the $SU(2)_{(A_1\bar{B}_1)}=SU(2)_G$ group rotating ${(A_1\bar{B}_1)}$. Therefore, it is natural to suggest that the open string is attached to the giant graviton D4-brane described by the following determinant 
operator,\footnote{Note that a different proposal for the giant corresponding to this operator was put forward in \cite{Berenstein:2008dc} as being two 
$\mathbb{CP}^2$ giants intersecting over a $\mathbb{CP}^1$, since the determinant splits into $\det A \det B$, but the $A$ and $B$ matrices are 
bifundamental, so it is not clear that they can be interpreted by themselves as D-branes, and we would instead suggest that there is an identification 
of those two would-be D-branes needed, leading to our interpretation. We would like to thank David Berestein for comments on his work.}
 \be\label{Giant2}
 {\cal O}_g=\ep_{m_1,...,m_N}\ep^{p_1,...,p_N}(A_1{B}_1)^{m_1}_{p_1}...(A_1{B}_1)^{m_N}_{p_N}.
\ee

The full bosonic symmetry of the above string vacuum is $U(1)'$, together with 
$SU(2)_G\times U(1)_D\times SO(3)_r$,  a subgroup of $SU(2|2)$, whose generators commute with $U(1)'$, 
where $SO(3)_r$ acts on the 3 worldvolume coordinates $y_i$, the generator $D=\Delta-J$.

We can now easily identify this symmetry with the pp wave isometry group.
The breaking of $SU(4)\rightarrow SU(2)_L\times SU(2)_R\times U(1)_u$ corresponds to $SU(4)\rightarrow SU(2)_G\times SU(2)_{G'}\times U(1)'$, 
meaning that $U(1)_u$ identified with $U(1)'$, $SU(2)_L$ with $SU(2)_G$ and $U(1)_D$ with $U(1)_\pm$. Then $SO(3)_r$ is the same in both cases, 
meaning the total symmetry is $U(1)'\times SU(2)_G\times U(1)_D\times SO(3)_r$ = $U(1)_u\times SU(2)_L\times U(1)_\pm\times SO(3)_r$.

The fact that ${\cal O}_g$ is not charged under the Casimir $J$ of $SU(2)_{G'}=SU(2)_{(A_2\bar{B}_2)}$, i.e,
\be\label{ju} 
J_{(A_2\bar{B}_2)}\big[(A_1{B}_1)\big]=0\,,
\ee
is understood as being due to the fact the propagating direction of the open string is parallel to the giant graviton D4-brane, hence the 
propagation direction for the string is different than the propagating direction for the giant, i.e. the charge $J$ for the open string is different than the 
charge $J_1=J_{(A_1\bar B_1)}=$ Casimir of $SU(2)_G$ for the giant, under which the giant has zero energy ($J_1(A_1)=J_1(B_1)=1/2$ and $\Delta=1/2$ give
$(\Delta-J_1)(A_1B_1)=0$).

The identification of $U(1)_D$ with $U(1)_\pm$ means that $J$ (the Casimir of $SU(2)_G$) is identified with $J_\chi$, the angular momentum in the 
$\chi$ direction of $\mathbb{CP}^3$, so that we can formally write 
\be\label{jotachi}
\mathbf{J}_{\chi}(A_2)=\mathbf{J}_{\chi}(B_2)=\frac{1}{2},\,~~~~~~\mathbf{J}_{\chi}(A_1)=\mathbf{J}_{\chi}(B_1)=0\,. 
\ee

From (\ref{jotachi}) we have 6 combinations in the adjoint of the first $U(N)$, classified according to 
$\Delta-J_{\chi}$ as  
\bea\label{building}
(\Delta-J_{\chi})&=&0:~~~~A_2 B_2,\nn\\
(\Delta-J_{\chi})&=&1/2:~ A_2B_1,\; A_2\bar{A}_1,\; A_1 B_2,\; \bar{B}_1 B_2,\nn\\
(\Delta-J_{\chi})&=&1: ~~~~A_1 B_1.\label{deltaminusj}
\eea
Summarizing, we would like to describe the vacuum of the open string-brane system by the following operator in ABJM
\be\label{FTvacuum}
\ep_{m_1,...,m_N}\ep^{p_1,...,p_N}(A_1B_1)^{m_1}_{p_1}...(A_1B_1)^{m_{N-1}}_{p_{N-1}}[W]^{m_N}_{p_N}\,.
\ee
It has been argued in \cite{Balasubramanian:2004nb, Berenstein:2006qk} that if we set an $(A_1B_1)$ at the border of $W$, the operator factorizes, 
so we do not want to consider that situation, although it should be interesting to study that phenomenon at both sides. In fact, an operator as 
(\ref{FTvacuum}) can be expanded in terms of traces (closed strings), but for maximal giant, which is the case we are 
considering, the mixing with closed strings is suppressed.  
This operator carries anomalous dimension minus $J_\chi$ charge of $\Delta-J_{\chi}=N-1$.

Note that for the string in the pp wave, we obtained a maximum $J_\chi$ of order $J_\chi\sim \sqrt{N}$. This suggests that $W$ in (\ref{FTvacuum}) should 
contain at most ${\cal O}(\sqrt{N})$ $(A_2B_2)$ combinations, though it is not clear why we should have this constraint from a field theory point of view. 
Perhaps for larger $J_\chi$ the open string oscillation starts to modify the giant graviton 
itself (which has an energy of $N-1$, as seen above), here assumed to be a fixed background.

As in the BMN case \cite{Berenstein:2002jq}, we should relate excitations of the string theory ground state 
with appropriate insertions into the string of operators $W$. 

Excitations in directions $X^M$ coming from the $AdS_4$, i.e. $y_i$,  correspond to insertion of covariant derivatives $D_i$ (with $\Delta-J_ {\chi}=1$)
in the dual operator, but we are not going to consider those here. 
Excitations along $\CP^3$ should be identified with insertions of composite operators. From (\ref{deltaminusj}) we find that the 
last excitation with frequency $\Delta-J_{\chi}=1$, corresponding to the direction $u$, is $A_1B_1$.  
A single excitation along the $(w_a,\bar w_a)$ directions ($X^N$, $N=5,...,8$) in the gravity side increases the energy of the vacuum by $1/2$, 
as we can see from (\ref{bosexc}). 
From (\ref{deltaminusj}) we see that the insertions increasing the energy of the ground state (\ref{FTvacuum}) by 1/2 are given by
\be
A_2B_1,\; A_2\bar{A}_1,\; A_1 B_2,\; \bar{B}_1 B_2\,.
\ee

For example, one of those excitations should be given by the insertion of an $(A_2B_1)$ i.e, corresponds to the operator,
\be\label{FT1ex}
 {\cal O}_l=\ep_{m_1,...,m_N}\ep^{p_1,...,p_N}(A_1B_1)^{m_1}_{p_1}...(A_1B_1)^{m_{N-1}}_{p_{N-1}}[(A_2B_2)^l(A_2B_1)(A_2B_2)^{J-l}]^{m_N}_{p_N}\,,
\ee
which has $\Delta-\mathbf{J}_{\chi}=\frac{1}{2}$. As for single trace operators, higher (massive) oscillator states of the open string, are described by operators with the insertions above accompanied by a phase position-dependent factor representing the given level. Explicitly, we associate the following operator with a single excitation of the string
\bea
{\cal O}_n&=& \ep_{m_1,...,m_N}\ep^{p_1,...,p_N}(A_1B_1)^{m_1}_{p_1}...(A_1B_1)^{m_{N-1}}_{p_{N-1}}\times\cr
&&\times\sum_{l=0}^{J}\left[(A_2B_2)^l(A_2B_1)(A_2B_2)^{J-l}\right]^{m_N}_{p_N}\cos\left(\frac{\pi n l}{J}\right).\label{FTOS}
\eea

\section{Anomalous dimension of ABJM operators}

\subsection{Single excitation}

We now move to the computation of the anomalous dimension of the operator in (\ref{FTOS}).
We show in Appendix B that the leading planar contribution comes only from interactions of the open chain, i.e. the term in square brackets. 
For single trace operators in the planar limit, at two-loops we only get mixing between nearest neighbours (from the point of view of the $(AB)$ pairs, 
i.e. next to nearest neighbours from the point of view of individual $A$ and $B$ fields), through the interactions terms in the Lagrangean
\bea
V&=&\frac{4\pi^2}{k^2}{\rm Tr}
\left[-2(\bar{B}_2\bar{A}_2\bar{B}_1B_1A_2B_2)+(\bar{B}_1\bar{A}_2\bar{B}_2B_1A_2B_2)+(\bar{B}_2\bar{A}_2\bar{B}_1B_2A_2B_1)\right]\cr
&=&\frac{4\pi^2}{k^2}{\rm Tr}\left[-2(\bar{B}_2\bar{A}_2\bar{B}_1B_1A_2B_2)+((\bar{B}_1\bar{A}_2)\bar{B}_2B_1(A_2B_2))
+((\bar{B}_2\bar{A}_2)\bar{B}_1B_2(A_2B_1))\right]
\,, \cr
&&\label{interac}
\eea
the first term is diagonal, the second one moves the impurity to the left (adding for free an inert $A_2$ on the left, this term connects 
$A_2B_1(A_2B_2)$ with $\bar A_2 \bar B_2 (\bar A_2\bar B_1)$)  and the third one move the impurity to the right
(adding an intert $A_2$ on the left, it connects $A_2B_2(A_2B_1)$ with $\bar A_2 \bar B_1(\bar A_2 \bar B_2)$). 

Then, just like in the BMN case for ${\cal N}=4$ SYM,  the operator ${\cal O}_n$ in (\ref{FTOS})
diagonalizes the interaction term (\ref{interac}) and the action of the interaction term on it produces a global phase 
(independent of the sum index $l$) coming from
\be\label{cosshift}
-2\cos\left(\frac{\pi n(l)}{J}\right)+ \cos\left(\frac{\pi n(l+1)}{J}\right)+\cos\left(\frac{\pi n(l-1)}{J}\right)=2\cos\left(
\frac{\pi n(l)}{J}\right)\left[-1+\cos\left(\frac{\pi n}{J}\right)\right].
\ee

Collecting the results from Appendix B, we can write the two-point function at one-loop as
\bea
\frac{\langle {\cal O}_n(x){\cal O}_n(0)\rangle_{1-{\rm loop}}}{\langle {\cal O}_n(x){\cal O}_n(0)\rangle_{{\rm tree}}}
&=&1+8\lambda^2\left[1-\cos\left(\frac{\pi n}{J}\right)\right]{\rm ln}(x\Lambda)\cr
&\equiv& \left(1+(\Delta-J)^{\rm anom.}\ln (x\Lambda)\right) \,.
\label{anomvacuum}
\eea

Note that at tree level (classical dimension) $\Delta=J_ {\chi}+1/2$ for our operator.
Expanding the cosine for small  $\frac{\pi n}{J}$ we get that the contribution to the anomalous dimension coming from the open string interactions is
\be\label{anomalous1}
(\Delta-J_{\chi})^{\rm anom.}=\frac{4\pi^2n^2}{J^2}\lambda^2\Rightarrow 
\Delta-J_{\chi}=\frac{1}{2}\left[1+\frac{8\pi^2n^2}{J^2}\lambda^2\right]. 
\ee
This agrees with the closed string calculation in (\ref{magnondisp}).

On the other hand, at strong coupling, the string theory result is given by the expression for the frequencies in the $\CP^3$ directions
in (\ref{bosexc}),  
\be 
w^{(r)}_n=\sqrt{\frac{1}{4}+\frac{n^2}{(\a' p^+)^2}}\,.
\ee 
After using $R^2/\a'=2^{5/2}\pi \sqrt{\lambda}$ and $p^+=J/R^2$ 
to get $\a' p^+=J/(2^{5/2}\pi\sqrt{\lambda})$, we obtain for the frequencies
\be\label{frecuencies}
 w^{(r)}_n=\frac{1}{2}\sqrt{1+\frac{2^3\pi^2n^2\lambda}{J^2}}.
\ee 
This agrees with the closed string result (\ref{magnondisp}).

Expanding for small $\lambda n^2/J^2$ so as to compare with the SYM case, we obtain
\be  
w^{(r)}_n\simeq\frac{1}{2}\left[1+\frac{4\pi^2n^2}{J^2}\lambda\right].
\ee

As in the closed string case,  it seems that the BMN scaling is violated for ABJM, since the strong coupling $\lambda\gg 1$  and weak coupling 
$\lambda\ll 1$ results have different behaviours. 
This is the same discrepancy from (\ref{magnondisp}) for the closed string case,  first noticed by \cite{Nishioka:2008gz, Gaiotto:2008cg}.

\section{Hamiltonian effective description on the ABJM operators}

In this section we would like to find a Hamiltonian description for the ABJM open string using Cuntz oscillators, similar to the ${\cal N}=4$ 
SYM case described in Appendix A. We will not give all the details which are the same as in the ${\cal N}=4$ SYM case, since they can be 
found in the Appendix.

The operator-state correspondence for 3 dimensional field theories relates operators on ${\mathbb R}^3$ with states on the cylinder $S^2\times {\mathbb R}_t$, 
found at the boundary of global $AdS_4$. Fields on ${\mathbb R}^3$ are KK reduced on $S^2$ and give creation and annihilation operators on ${\mathbb R}_t$, 
which can act on states. Fields without derivatives on ${\mathbb R}^3$  correspond to the zero modes of the KK reduction on $S^2$.
The new feature now is that we have pairs of fields that appear naturally, e.g. $(A_1B_1)(x)$, $(A_2B_1)(x)$, $(A_2B_2)(x)$, so we can consider them 
together under dimensional reduction on $S^2$. We denote $(A_2B_2)(x)\leftrightarrow a^\dagger$, $(A_2B_1)(x)\leftrightarrow b^\dagger$.
Like in the case of ${\cal N}=4$ SYM, the vacuum $|0\rangle_J$ for the open string is defined by acting with $J$ field objects on the true vacuum $|0\rangle$, 
but unlike that case, now we act with composite operators $(A_2B_1)(x)$ and $(A_2B_2)(x)$\footnote{For the sake of simplicity, we are going to drop the determinant factor along this section, but it is understood that the free indices $m_N, p_N$ are attached to it as in (\ref{FTOS}).}
\bea
&&[0_J]^{m_N}_{p_N}=[(A_2B_2)^J]^{m_N}_{p_N}\Rightarrow\cr
&&|[0]^{m_N}_{p_N}\rangle_J=[(a^\dagger)^J]^{m_N}_{p_N}|0\rangle.
\eea
But as explained in Appendix A, the oscillators appearing here are actually Cuntz oscillators, satisfying (\ref{cuntzosc}), because there is an implicit 
group structure (the creation operators have matrix indices) that means that the order matters, and then the Hilbert space can be mapped to the Hilbert 
space of Cuntz oscillators.

For excited states, we have also operators insertions of $(A_2B_1)(x)$ on ${\mathbb R}^3$, which we can replace by insertions of $b^\dagger$ along the 
string of $a^\dagger$'s, at site $i$. Equivalently, we can consider the {\em independent} Cuntz oscillators at each site $b_j$, as in (\ref{cuntzbi}).
In order for this to be a good definition, we need to have very few cases where the $b^\dagger$'s appear at the same site, namely we need to be in 
the "dilute gas" approximation.

The action of the interaction potential on operators through Feynman diagrams in ${\mathbb R}^3$ generates a Hamiltonian action on states
in the $S^2\times {\mathbb R}_t$ picture.
Indeed, by acting with the interaction potential $(\ref{interac})$ through Wick contractions onto a one-impurity operator
\be
[{\cal O}_{l}]^{m_N}_{p_N}=\left[(A_2B_2)^l(A_2B_1)(A_2B_2)^{J-l}\right]^{m_N}_{p_N}\,,\ee
we obtain the action (we have a factor of $4\pi^2/k^2$ in front of the action and a factor of $N^2$ coming from the index loops, 
instead of the factors $g^2_{YM}N/2$ for ${\cal N}=4$ SYM in the Appendix, allowing us 
to write the result in terms of the 't Hooft coupling $\lambda=N/k$)
\be
V\cdot[{\cal O}_{l}]^{m_N}_{p_N}=\lambda^2[-2{\cal O}_{l}+{\cal O}_{l+1}+{\cal O}_{l-1}]^{m_N}_{p_N}+{\rm 3-impurity} \;,
\ee
which is the fact we have actually used in (\ref{cosshift}). Therefore we can define the action of $V$ on the states in the dilute gas approximation 
(where various $b_j^\dagger$ excitations don't interact with each other) by the Hamiltonian terms in $V$:
\be 
\lambda^2\left[-b_l^{\dagger}b_l-b_l b^\dagger_l
+b_l^{\dagger}b_{l+1}+b_{l+1}^{\dagger}b_{l}\right]\;,
\ee 
plus terms with two $b^\dagger$'s and with two $b$'s. In fact, as in Appendix A, these terms are needed because of 1+1 dimensional relativistic invariance, 
requiring that we obtain the combination (field) $\phi_j=(b_j+b^\dagger_j)/\sqrt{2}$, which determines uniquely the interaction Hamiltonian.

The final result for the full interacting Hamiltonian, including the kinetic terms, is the nearest-neighbour result
\be 
H=\sum_{l=1}^J\frac{b_l^{\dagger}b_l+b_lb_l^{\dagger}}{2}+\lambda^2\sum_{l=1}^J\left(b_{l+1}^{\dagger}+b_{l+1}-b_{l}^{\dagger}-b_{l}\right)^2. 
\ee

As in Appendix A, we can do a Fourier transform from the oscillators $b_j$ to oscillators $b_n$ by
\be 
b_j=\sum_{j=1}^J\e^{\frac{2\pi i\,j\,n}{J}}b_n\,, 
\ee
after which the Hamiltonian becomes
\bea 
H&=&\sum_{j=1}^J\frac{b_nb_n^{\dagger}+b_n^{\dagger}b_n}{2}\cr
&&+\lambda^2 \sum_{j=1}^J\left[b_nb_n(\e^{ip}-1)+b_nb^{\dagger}_n(\e^{ip}-1)+b^{\dagger}_nb_n(\e^{-ip}-1)
+b^{\dagger}_nb^{\dagger}_n(\e^{ip}-1)\right]\,,\nn\\
\eea
where $p= 2\pi\,\,n/J$ is the magnon momentum. We further redefine
\be 
b_n=\frac{c_{n,1}+c_{n,2}}{\sqrt{2}}\,,\quad  b_{J-n+1}=\frac{c_{n,1}-c_{n,2}}{\sqrt{2}}\;,
\ee 
such that the Hamiltonian becomes
\bea\label{cshamil} &&H=\sum_{j=1}^{J/2}\frac{c_{n,1}c_{n,1}^{\dagger}+c_{n,1}^{\dagger}c_{n,1}}{2}+\frac{c_{n,2}c_{n,2}^{\dagger}+c_{n,2}^{\dagger}c_{n,2}}{2}\nn\\
&+&\lambda^2 \sum_{j=1}^J\left[\alpha_n (c_{n,1}+c_{n,1}^{\dagger})^2)-\alpha_n (c_{n,2}-c_{n,2}^{\dagger})^2-\beta_n[(c_{n,1}-c_{n,1}^{\dagger}),(c_{n,2}+c_{n,2}^{\dagger})]\right].
\eea
Here
\be 
\alpha_n=2(\cos(p)-1)=-4\sin^2\left(\frac{\pi n}{J}\right),\quad \beta_n=\sin\left(\frac{2\pi n}{J}\right).
\ee
In the dilute gas approximation (see Appendix A for more details)
we can see that the operators $b_n$ satisfy to leading order the usual harmonic oscillator algebra,  
\be 
[b_n,b_m^{\dagger}]\sim\delta_{m,n}+{\cal O}(1/J)\,, 
\ee 
from which it is easy to see that the last commutator in (\ref{cshamil}) vanishes.
Then the hamiltonian is given in the large $J$ limit by a sum of perturbed harmonic oscillators 
as in (\ref{generic}), and we can follow the same recipe as in Appendix A to do a Bogoliubov transformation on the 
Hamiltonian to find the eigenstates and their energy,
\be
\omega_n=\sqrt{1+4\lambda^2|\a_n|}=\sqrt{1+16\lambda^2\sin^2\left(\frac{p}{2}\right)}.
\ee
This agrees with the closed string result (\ref{magnondisp}) at weak coupling, 
and by expanding in $n/J\ll 1$ and in $\lambda\ll 1$ agrees also with the two-loop computation in section 5.  
But the result obtained in here is much more general, since it resums the 2-loop Hamiltonian contributions to the energy, 
and it applies for arbitrary magnon momentum $p$, not just $p\ll 1$.

\section{Conclusions}

In this paper we have considered open strings ending on D4-brane giant gravitons in $AdS_4\times \mathbb{CP}^3$, and the operators dual to them in 
ABJM. The D4-brane giant moves in a $\mathbb{CP}^2\subset \CP^3$, and we considered a large R-charge limit for the operators corresponding to a pp wave limit 
in the gravity dual. We have described the operators corresponding to open strings with excitations on them, and the resulting magnon dispersion relation
coincides with the one in the closed string case, (\ref{magnondisp}). In particular, we calculated the first, two-loop, correction to the energy and 
noted the different scaling from the string theory result, as in the closed string case. We showed explicitly how to derive the magnon dispersion relation 
in the ${\cal N}=4$ SYM case using a Hamiltonian description based on Cuntz oscillators, that was only implicit in \cite{Berenstein:2002jq}, and then showed 
how to parallel that analysis for open string operators in ABJM. 

\section*{Acknowledgements}

The work of HN is supported in part by CNPq grant 301219/2010-9 and FAPESP grant 2013/14152-7, and the work of 
CC is supported in part by CNPq grant 160022/2012-6. We would like to thank Jeff Murugan for discussions and for comments on the 
manuscript, and to David Berenstein and Andrea Prinsloo for comments on the first version of this paper.
CC is grateful to Diego Correa for useful discussions.

\appendix

\section{Hamiltonian description for large charge ${\cal N}=4$ SYM operators using Cuntz oscillators}

In this Appendix we review the Hamiltonian calculation using Cuntz oscillators for the ${\cal N}=4$ SYM case, which was implicit in 
\cite{Berenstein:2002jq}. The result gives explicitly the $\sin^2(p/2)$ factor inside the square root for the energy, generalizing a bit 
the result in \cite{Berenstein:2002jq}.

As explained in \cite{Berenstein:2002jq}, one considers ${\cal N}=4$
SYM KK reduced on $S^3\times {\mathbb R}_t$
(the boundary of $AdS_5\times S_5$), and after the dimensional reduction
on the $S^3$ factor one gets a Hamiltonian description for SYM.
 
The SYM fields on ${\mathbb R}^4$ are organized in terms of $\Delta -J$, corresponding
to energy in the dual pp wave string theory. Here $J$ is a $U(1)$ R-charge that rotates the 
complex field $Z=X^1+iX^2$ by a phase. The vacuum is made up of $Z$ fields, with 
$\Delta-J=0$. The string oscillators are the fields with $\Delta-J=1$, namely
the 4 $\phi^I$'s, 4 derivatives of $Z$, $D_mZ$, and 8 fermions $\chi^a_
{J=1/2}$. Then there are fields of $\Delta -Z>1$, like $\bar{Z}$, $\chi^a_{J=-1/2}$
and the higher derivatives of the string oscillator fields. 

Under dimensional reduction on $S_3$, arising from the operator-state 
correspondence for conformal field theories (which in 4 dimensions relates $S_3\times {\mathbb R}_t\leftrightarrow {\mathbb R}^4$, 
the same way as in the more familiar 2 dimensional case it relates the cylinder $S^1\times {\mathbb R}_t$ with the plane ${\mathbb R}^2$),
the KK modes of a field, which correspond
to higher spherical harmonics on the sphere $S^3$, are mapped to the higher 
derivative modes $D_{m_1}...D_{m_n}$ of ${\mathbb R}^4$ fields. The ${\mathbb R}_t$ ''mass''
under $S^3$ KK reduction (i.e., frequency of harmonic oscillators) 
has an extra term due to the curvature coupling to the scalars, so that 
even the constant mode of $Z$, i.e. the field $Z$ on ${\mathbb R}^4$, has frequency equal to
1. The corresponding harmonic oscillators are called ${(a^\dagger)^i}_j$ (here 
$i,j$ are $SU(N)$ indices). The ${\mathbb R}^4$ fields of $\Delta-J=1$, $\phi^I, 
D_mZ$, correspond to their constant mode on $S_3$, having frequency equal to 2, and the corresponding 
oscillators are denoted by ${(b^\dagger)^i}_j$ (here we have supressed 
the $I,m$ indices on $b$). Together, the set of $a^\dagger$, $b^\dagger$ and higher 
KK modes are called $a^\dagger_{\alpha}$. 

A large $J$ charge single trace operator (such an operator is leading in the large $N$ limit)
then corresponds on ${\mathbb R}_t$ to a ordered
string (or ''word'') of $a^\dagger_{\alpha}$'s acting on the vacuum $|0\rangle$. 
In the sector of fixed $J$ (corresponding to fixed $p^+$ momemntum on a 
pp wave string), the vacuum $|0\rangle_J$, is $\Tr[(a^\dagger)^J]|0>$, mapped to the 
operator $\Tr[Z^J]$ on ${\mathbb R}^4$. We will drop the $J$ index in the following, 
assuming we are in the $J$ vacuum. 

Next one uses the observation of \cite{Gopakumar:1994iq}, or rather of \cite{VDN}, that 
the Hilbert space of $n$ independent large $N$ random matrices acting on a vacuum, 
$\hat{M}_1\hat{M}_2\hat{M}_3...|0\rangle$ (where the order is important, so 
that one has to consider a {\em word} made of $\hat{M}_i$'s) is the 
same as the Hilbert space of so-called Cuntz oscillators $a_i$, 
$i=1,...,n$, satisfying 
\be
a_i|0\rangle=0;\;\;\; a_ia_j^\dagger=\delta_{ij};\;\;\; \sum_{i=1}^n a_i^\dagger a_i=1-|0\rangle\langle0|\label{cuntzosc}
\ee
and no other relations (in particular no relations between $a^\dagger_ia^\dagger_j$
and $a^\dagger_ja^\dagger_i$, so that the order is important). 

For a single Cuntz oscillator we would have 
\be
a|0\rangle=0;\;\;\; aa^\dagger=1;\;\;\; a^\dagger a=1-|0\rangle\langle0|\;,
\ee
and the number operator is 
\be
\hat{N}=\frac{a^\dagger a}{1-a^\dagger a}=\sum_{k=1}^{\infty} (a^\dagger)^ka^k.
\ee

In general,
\be
\hat{N}=\sum_{k=1}^{\infty}\sum_{i_1,...i_k}a^\dagger_{i_1}...a^\dagger_{i_k}a_{i_k}
...a_{i_1}.
\ee

Note that the equality of the large $N$ Matrix Hilbert space with the Cuntz algebra Hilbert space
is a fact and is independent of the main subject of \cite{Gopakumar:1994iq},
which was to describe random matrix correlators by a ''master field''
$M(z)$ (or $\hat{M}(a, a^\dagger)$), i.e. such that 
\be
\Tr [M^p]=\lim_{N\rightarrow \infty} \int {\cal D} M e^{-N \Tr\; V(M)}
\frac{1}{N} \Tr[M^p]=\langle 0|\hat{M}(a, a^\dagger)^p|0\rangle.
\ee
For that, one needed to define inner products and use properties of Matrix 
models. Here we restrict ourselves to Hamiltonians acting on states, 
for which \cite{VDN} suffices.

One thing which is not taken into account in the formalism is the fact 
that our Hilbert space is for traces of matrices, which are cyclic, thus 
cyclicity must be imposed by hand.

{\bf Hamiltonian}

The particular case studied in \cite{Berenstein:2002jq} 
involves particular types of words, with mostly $Z$'s, corresponding to $a^\dagger$'s, and few $b^\dagger$'s
(the ''dilute gas'' approximation). Thus an additional simplification 
was used: in a $a^\dagger...a^\dagger b^\dagger a^\dagger... a^\dagger b^\dagger a^\dagger...a^\dagger|0\rangle$ type string of length
$J$ (with $J$ $a^\dagger$'s), we can forget the $a$ Cuntz oscillators and consider that we have 
a chain of $J$ sites, and at each site a different {\em independent} Cuntz oscillator
$b_j^\dagger $  (here $j=1,...,J$ labels sites), i.e. consider
\be
[b_i, b_j]=[b_i^{\dagger}, b_j]=[b_i^{\dagger}, b_j^{\dagger}]=0 ,\;\;\; i\neq j\;,
\ee
and
\be
b_ib_i^{\dagger} =1, \;\;\; b_i^{\dagger} b_i=1-(|0\rangle\langle 0|)_i;\;\; b_i|0\rangle_i=0.\label{cuntzbi}
\ee

Of course, there is still the supressed index $I,m$ corresponding to the 
type of oscillator. Defining Fourier modes,
\be
b_j=\frac{1}{\sqrt{J}}\sum_{n=1}^Je^{\frac{2\pi i jn}{J}} b_n\;,
\ee
we get the commutation relations
\be
[b_n, b_m^{\dag}]=\frac{1}{J}\sum_{j=1}^Je^{\frac{2\pi ij(m-n)}{J}}
(|0\rangle\langle 0|)_j;\;\;\; [b_n,b_m]=[b_n^\dagger,b_m^\dagger]=0.
\ee

This is in general a complicated operator, but if we act on states 
in the dilute gas approximation, i.e. on states
\be
|\psi_{\{ n_i\} }\rangle=|0\rangle_1...|n_{i_1}\rangle...|n_{i_k}\rangle...|0\rangle_J\;,
\ee
we get 
\be
[b_n, b_m^{\dagger}]|\psi_{\{ n_i\} }\rangle=\left(\delta_{nm} -\frac{1}{J}
\sum_k e^{2\pi i i_k\frac{m-n}{J}} \right)|\psi_{ \{ n_i \} }\rangle\;,
\ee
that is, the operator gives $1/J$ corrections in the dilute gas 
approximation. In particular, $[b_n,b_m^\dagger]|0\rangle=\delta_{m,n}$, as for 
usual oscillators. Since also $b_n|0\rangle=0$ as we can easily check, the 
$b_n$' act as usual creation/annihilation operators, exactly on the vacuum 
and approximately on dilute gas states.

With these Cuntz oscillators for the b's, the interaction term 
\be
L_{\rm int}=-\frac{g_{YM}^2}{2}\Tr [z, \phi] [\bar{z}, \phi]\;,
\ee
where $4\pi g_s=g_{YM}^2$, in the lagrangian becomes equivalent to
\be
-\frac{g_sN}{\pi}\sum_j  (\phi_j -\phi_{j+1})^2\;,
\ee
where $\phi_j=(b_j+b_j^{\dagger})/\sqrt{2}$. (With the usual 
oscillator one would have extra factors of $1/\sqrt{2}$ in the 
$(b_j^{\dagger})^2$ terms.) An obvious term is the one with 2 $b^{\dagger}$'s,
$\sum_j b^{\dagger}_j b^{\dagger}_{j+1}+b^{\dagger}_{j+1}b^{\dagger}_j-(b^{\dagger}_j)^2
-(b^{\dagger}_{j+1})^2 $. It arises from the contraction of the $\bar{z}$ in 
$L_{int}\sim 2\bar{z}\phi z \phi -\bar{z}\phi^2 z-\bar{z}z\phi^2$ with 
one $z$ in the state. All the other terms can be obtained similarly, or we can consider the fact that
1+1 dimensional relativistic invariance requires that the combination $\phi_j=(b_j+b_j^\dagger)/\sqrt{2}$ to appear in the 
interaction Lagrangean. We will denote by $ \lambda=g_{YM}^2N$ the 't Hooft coupling.

Then the total hamiltonian (equal to a free part plus the interaction 
part, that is, minus the interaction Lagrangean from above) should be 
\be
H=\sum _{j=1}^J\frac{b_jb_j^{\dagger}+b_j^{\dagger} b_j}{2} +
\frac{\lambda}{8\pi^2}\sum_{j=1}^J[(b_{j+1}+b_{j+1}^{\dagger})(b_j+b_j^{\dagger})
-(b_j+ b_j^{\dagger})^2].
\ee

Notice however that $H|0\rangle=const.|0\rangle+(\lambda/8\pi^2) \sum_j (b_{j+1}^\dagger b_j^\dagger-b_j^\dagger b_j^\dagger)
|0\rangle\neq 0$. Since we know that $\Tr Z^J$ is a good vacuum ($\Delta -J$ 
remains zero, as this state is BPS), it follows that
on ${\mathbb R}_t$ we must have $H|0\rangle=0$, hence we must assume 
that susy cures the discrepancy e.g. by fermion loops, and one can have a redefined 
Hamiltonian $\tilde{H}$ such that $\tilde{H}|0\rangle=0$. We will thus put 
$H|0\rangle=0$ in the following by hand.

After going to the Fourier modes $b_n$ and then redefining the oscillators
by
\bea
b_{n}&=&\frac{c_{n,1}+c_{n,2}}{\sqrt{2}}\nonumber\\
b_{J-n}&=&\frac{c_{n,1}-c_{n,2}}{\sqrt{2}}\;,
\eea
the hamiltonian becomes (for $J=2k+1$, term $n=0 $, or rather $J$,  drops out
of the sum from 0 to $J$)
\bea
H&=&\sum_{n=1}^{[J/2]}\left[\frac{c_{n,1}^{\dag}c_{n,1}+c_{n,1}c_{n,1}^{\dagger}}{
2}+\frac{c_{n,2}^{\dagger}c_{n,2}+c_{n,2}c_{n,2}^{\dagger}}{2}+\right.
\nonumber\\
&&\left.\alpha_n(c_{n,1}+c_{n,1}^{\dagger})^2-\alpha_n(c_{n,2}-c_{n,2}^{\dagger})^2
+\beta_n[c_{n,1}-c_{n,1}^{\dagger}, c_{n,2}+c_{n,2}^{\dagger}] \right]\;,
\eea
where
\bea
\alpha_n&=&\frac{\lambda}{8\pi^2} (\cos (2\pi n/J)-1)= -\frac{\lambda}{(2\pi)^2} \sin ^2\frac{\pi n}{J}  
\nonumber\\
\beta_n&=& i \frac{\lambda}{(2\pi)^2} \sin (2\pi n/J).
\eea
However, with our commutation relation for $[b_n, b_m^{\dagger}]$ one 
can check that the commutator term vanishes and the hamiltonian is 
now diagonal, albeit with nontrivial oscillators $c_{n,a}$.

Moreover, it is now exactly in the form of a sum 
of perturbed oscillators. Indeed, a generic hamiltonian 
\be
H=\frac{aa^{\dagger}+a^{\dagger} a}{2}\pm \frac{\mu^2}{2}\frac{(a\pm a^{\dagger})^2}{2}
=\left(1+\frac{\mu^2}{2}\right)\frac{aa^{\dagger}+a^{\dagger}a}{2}\pm \frac{\mu^2}{4}
(a^2+{a^{\dagger}}^2)\;,\label{generic}
\ee
under the Bogoliubov transformation 
\bea
b&=&\alpha a \pm \beta a^{\dagger}\nonumber\\
\alpha -\beta &=& 1/\sqrt{\omega}\;\;\; \alpha +\beta =\sqrt{\omega}
\nonumber\\
\omega&=&\sqrt{1+\mu^2}
\eea
becomes
\be
H=\omega\frac{bb^{\dagger}+b^{\dagger}b}{2}.
\ee

Here if we had usual oscillators we would have 
$[a,a^{\dagger}]=[b,b^{\dagger}]$, i.e. the commutation relations would be
preserved. In the case of a single Cuntz oscillator this is still true, 
but now for many different Cuntz oscillators the algebra will change.

A Bogoliubov transformation in terms of usual oscillators will give 
a new vacuum after the transformation, since $a|0\rangle=0$ will imply 
$b|0\rangle\neq 0$, and $b|0'\rangle=0$ gives $|0'\rangle=\exp(-\beta (a^\dagger)^2/\alpha)|0\rangle$. 
But now we don't have usual oscillators.

Applying this Bogoliubov transformation to our Hamiltonian we get 
\be
H=\sum_{n=1}^{J/2}\omega_n\left[\frac{\tilde{c}_{n,1}^{\dagger}\tilde{c}_{n,1}+
\tilde{c}_{n,1}\tilde{c}_{n,1}^{\dagger}}{
2}+\frac{\tilde{c}_{n,2}^{\dagger}\tilde{c}_{n,2}+\tilde{c}_{n,2}\tilde{
c}_{n,2}^{\dagger}}{2}\right]\;,
\label{hami}
\ee
where the relations between oscillators are
\bea
&& \tilde{c}_{n,1}=a_nc_{n_1}+b_nc_{n,1}^\dagger\nonumber\\
&&\tilde{c}_{n,2}=a_nc_{n_1}-b_nc_{n,1}^\dagger\nonumber\\
&&a_n=\frac{(1+\alpha_n)^{1/4}+(1+\alpha_n)^{-1/4}}{2}\nonumber\\
&&b_n=\frac{(1+\alpha_n)^{1/4}-(1+\alpha_n)^{-1/4}}{2}\;,
\eea
and the energy of the eigenstates is
\be
\omega_n=\sqrt{1+4|\alpha_n|}=\sqrt{1+ \frac{4\lambda}{(2\pi)^2} \sin^2 \frac{\pi n}{J}}
=\sqrt{1+\frac{4g_sN}{\pi} \sin^2 \frac{\pi n}{J}}.
\ee
As we can see, this calculation was
exact, both in $\lambda$ and in $n/J$, as long as $J\rightarrow\infty$ and we have a dilute gas approximation.
For $n\sim 1\ll J$, we obtain for the energy
\be
\omega_n\simeq \sqrt{1+\frac{\lambda  n^2}{J^2}}=\sqrt{1+\frac{4\pi g_sN  n^2}{J^2}}\;,
\ee
which is the case used in \cite{Berenstein:2002jq}. 

Note that the result here is exact in $\lambda$, but a priori needed not be, since the calculation was one-loop in SYM, and the square root form came 
about because of the Bogoliubov transformation, so was a sort of resumming of various one-loop contributions, similar to the exponentiation of the IR
divergences of gluon amplitudes in the Sudakov factor $\sim \exp[\lambda a_1+{\cal O}(\lambda^2)]$. 

That means that in general, we expect the $\lambda$ inside the square root to be replaced by a more general function of $\lambda$, and this 
is indeed what happens in the ABJM case.

\section{First correction to anomalous dimension} \label{A1}

In this appendix we compute the anomalous dimension of the first excited state (\ref{FTOS}) at first order in $\lambda^2=16\pi^2N^2/k^2$.
The basic propagators we need are
\be\label{basicprog}
\langle A^i_{a\bar{b}}(x)\bar{A}^j_{\bar{c}d}(0) \rangle=\langle B^i_{\bar{b}a}(x)\bar{B}^j_{d\bar{c}}(0) \rangle=\frac{\delta^{ij}\delta_{ad}\delta_{\bar{b}\bar{c}}}{4\pi|x|} \,,
\ee 
from where obtain the following composite operators two-point functions,
\be\label{comp}
 \langle (A^iB^i)_{ab}(x)\overline{(A^jB^j)}_{cd}(0) \rangle=N\frac{\delta^{ij}\delta_{ad}\delta_{bc}}{16\pi^2|x|^2}\,.
\ee 
In our conventions the scalar potential in ABJM is given by
\be
     V=\Tr\left(|M^{\alpha}|^2+|N^{\alpha}|^2\right),
\ee
where
\bea
  M^{\alpha}&=&\frac{2\pi}{k}\Big(2\bar{B}^{[\alpha}B_{\beta}\bar{B}^{\beta]}+A^{\beta}  
  \bar{A}_{\beta}\bar{B}^{\alpha}-\bar{B}^{\alpha}\bar{A}_{\beta}A^{\beta}
  +2\bar{B}^{\beta}\bar{A}_{\beta}A^{\alpha}-2A^{\alpha}\bar{A}_{\beta}\bar{B}^{\beta}\Big),\nonumber\\
  N^{\alpha} &=& \frac{2\pi}{k}\Big(2A^{[\alpha}\bar{A}_{\beta}A^{\beta]}+\bar{B}^{\beta}  
  B_{\beta}A^{\alpha}-A^{\alpha}B_{\beta}\bar{B}^{\beta}
  +2A^{\beta}B_{\beta}\bar{B}^{\alpha}-2\bar{B}^{\alpha}B_{\beta}A^{\beta}\Big).\cr
  &&\label{mandn}
\eea

Since the potential is purely sextic (thus proportional to $\lambda^2$, i.e. $g_{YM}^4$), the first quantum corrections to the anomalous dimension
appear at two-loops (we can also check that the 6-vertex connecting 3 fields in one operator with 3 fields in another gives a 2-loop graph).
Since the composite operators we are going to use are in the adjoint of $U(N)$, the computation is very similar as the open string on giant 
in the ${\cal N}=4$ SYM \cite{Balasubramanian:2002sa}.
We split the computation as follows: 

\begin{itemize}
\item {\it Tree level}: The three level two point function of the state (\ref{FTOS}) is given by
\be
 \langle {\cal O}_n(x){\cal O}_n(0)\rangle_{tree}=N^{N+J-1}\frac{N!^2\,(N-1)!}{(4\pi|x|)^{2(N+J-1)}}\,.
\ee
As we see from (\ref{comp}), each propagator for composite fields $(AB)$ contributes with a factor of $N$, producing a total contribution of $N^{N+J-1}$. The factor of $(N-1)!$ counts all the possible contractions between $A_1B_1'$s, and $N!^2$ comes from the full contractions of two pairs of Levi-Civita symbols. $N^{J-1}$ comes from planar delta's contractions in the chain of $(A_2B_1)'s$. 

\item{\it\bf{Two-loops.}}

\item {\it The contribution coming from the interaction between giant graviton bits (components) vanishes}:

Since the giant graviton dual is built only from $(A_1B_1)$ composites, the only (possibly) non-vanishing contributions from interactions are those with only $A_1$ and $B_1$'s in it. 
\bea\label{ab1}
  |M^{1}|^2&=&\frac{4\pi^2}{k^2}\Big(-A_{1}  
  \bar{A}_{1}\bar{B}_{1}
  +\bar{B}_{1}\bar{A}_{1}A_{1}\Big)\Big(-A_{1}  
  \bar{A}_{1}\bar{B}_{1}
  +\bar{B}_{1}\bar{A}_{1}A_{1}\Big)^{\dagger},\nn\\
  &=&\frac{4\pi^2}{k^2}\Big(A_{1}\bar{A}_{1}\bar{B}_{1}\bar{A}_{1}{A}_{1}{B}_{1}
  -\bar{B}_{1}\bar{A}_{1}A_{1}\bar{A}_{1}  
  {A}_{1}{B}_{1}\nn\\
  &&~~~~~-A_{1} 
  \bar{A}_{1}\bar{B}_{1}{B}_{1}{A}_{1}\bar{A}_{1}
  +\bar{B}_{1}\bar{A}_{1}A_{1}{B}_{1}{A}_{1}\bar{A}_{1}\Big)  
\eea
One can see that there are
\be
 \left( \begin{array}{c}
2N-2\\
3\\
\end{array} \right)^2(2N-2)!
\ee
graphs of the following form for each term in (\ref{ab1}),
\begin{center}
\includegraphics[scale=0.5]{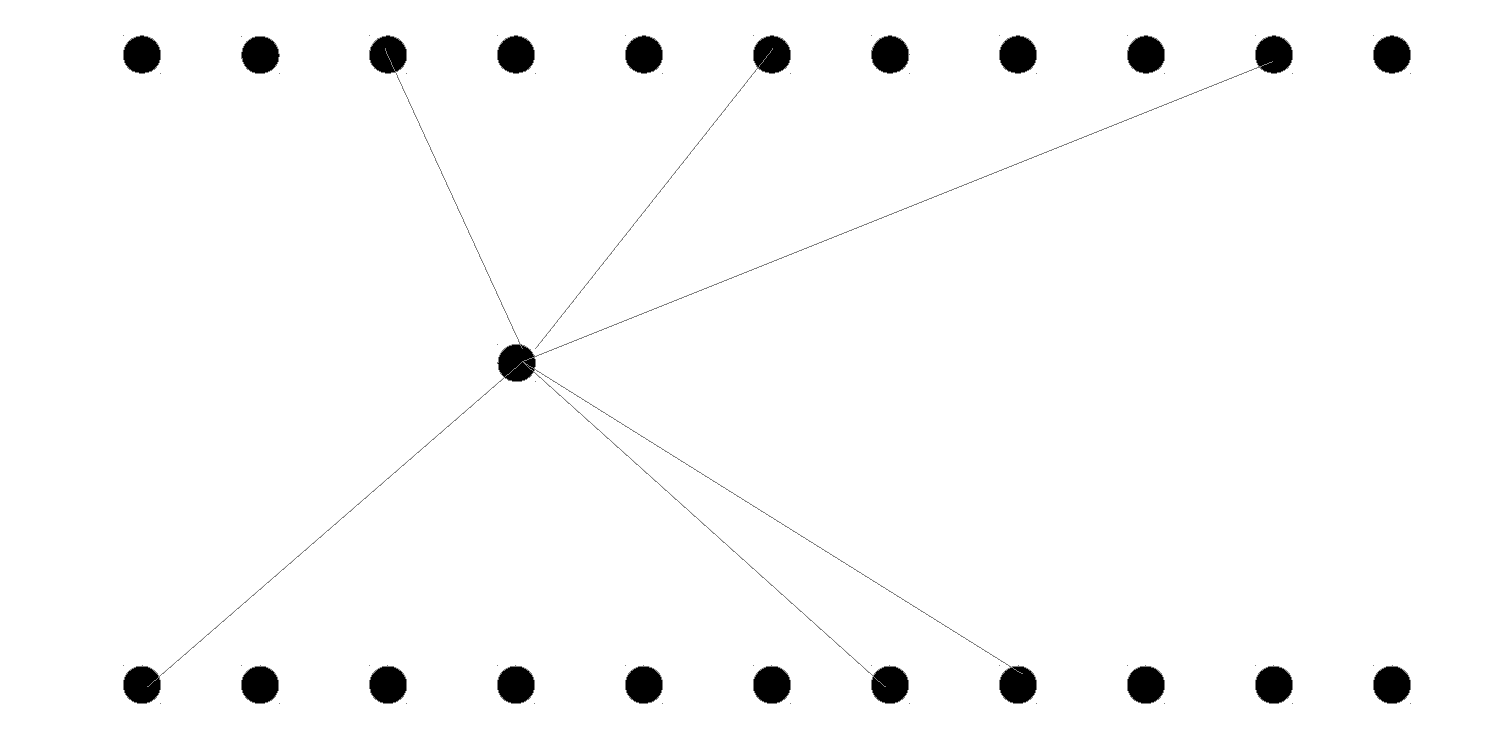}
\end{center}
but for each of these graphs, there exist 16 different ways of contracting the fields into the vertex with the fields in the operators. 
(For example, for the  first term in (\ref{ab1}) $A_{1}\bar{A}_{1}\bar{B}_{1}\bar{A}_{1}{A}_{1}{B}_{1}$, there are two ways of contracting $A_1$ with it, times two ways of contracting $\bar{A}$ with it, times four ways of contracting the remaining $B'$s and $\bar{B}'$s.)

Each of those 16 different contractions produces 8 odd permutations, plus 8 even permutations. Therefore, since each term in (\ref{ab1}) contributes 
the same number of odd permutations plus the same number of even permutations, the total sum just vanishes. 

It is interesting to note that in this case the giant graviton does not develop anomalous dimension (at least at one-loop) without the need for supersymmetry.

\item {\it Contractions between open strings bits}: Contractions between giant gravitons bits are as at tree level, $N!^2(N-1)!$,  as well as $N^{N}$ 
coming from propagators. Into any planar diagram in our conventions, each closed line contributes with an $N$, each propagator contributes with one (\ref{basicprog}), and each vertex contributes with a $4\pi^2/k^2$. Hence at tree level for example,
\begin{center}
\includegraphics[scale=0.5]{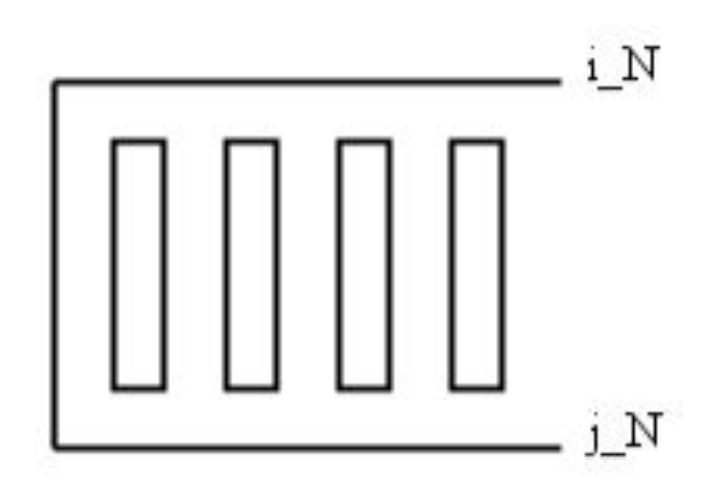}
\end{center}
we have from the open string $N^{J-1}\delta_{i_N,j_N}$, as we already mentioned. 
The first leading planar correction for the open string chain looks like
\begin{center}
\includegraphics[scale=0.5]{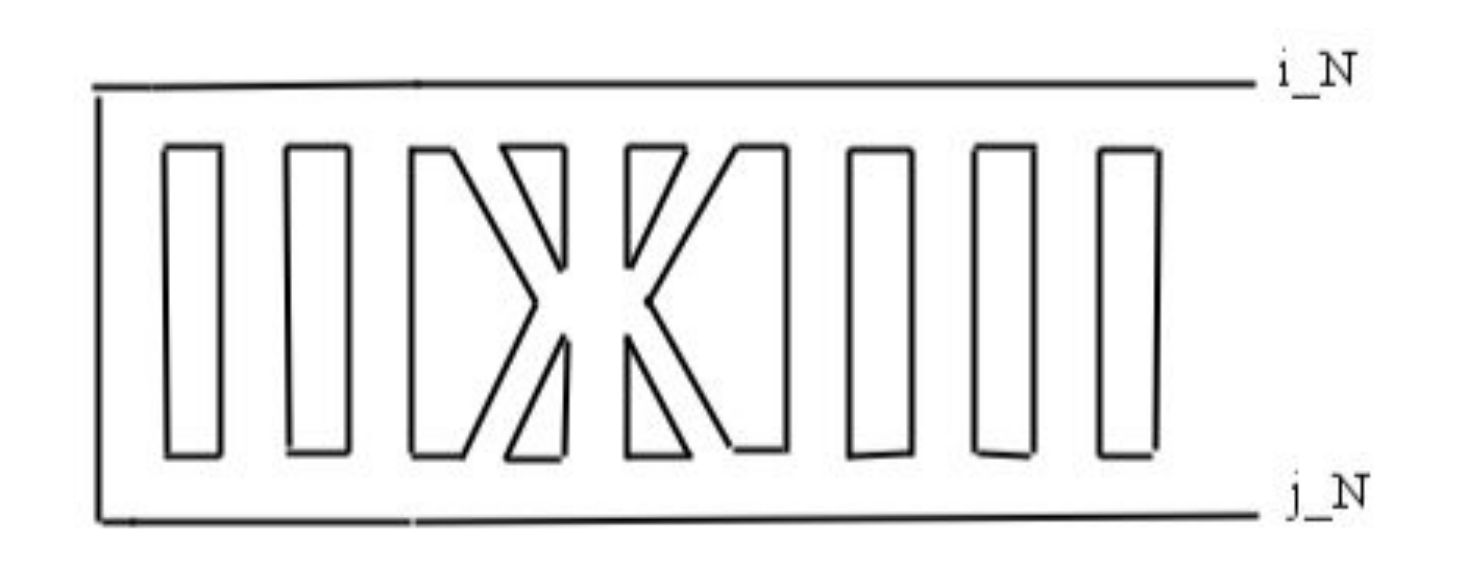}
\end{center}
and produces a coupling constant and $N$ dependence from the open string part of the operator of $N^{J-3}\frac{4^2\pi^2N^4}{k^2}=4^2\pi^2N^{J-1}\lambda^2$ .

The total result for the diagram is
\be
\left(\frac{N}{16\pi^2|x|^2}\right)^{N+J-1}4^2\pi^2 N!^2(N-1)!\,\lambda^2\, {\cal I}(|x|)= 4^2\pi^2\lambda^2\,\langle {\cal O}_n(x){\cal O}_n(0)\rangle_{tree}\, {\cal I}(|x|)\;,\label{O-O}
\ee
where (there are two equal divergent contributions to the result, one at $y=0$ and another at $y=x$)
\be
 {\cal I}(|x|)=\frac{|x|^3}{(4\pi)^3}\int\frac{d^3 y}{|x-y|^3|y|^3}\sim\frac{1}{8\pi^2}{\rm ln}(x\Lambda)\,.
\ee
Here we have introduced a cut-off $\Lambda$ in order to regulate  de divergent behaviour of the integral.

\item {\it Contractions between operators in the open strings and the giant}: From the giant graviton bits, we have $(N-1)^2$ possible choices 
for the fields that will interact with open string. That leaves $(N-2)!$ ways to contract the remaining bits freely and produces an 
additional $(N-2)!^2$ coming form the contractions of the Levi-Civitas. Again we obtain an $N^N$ factor from the propagators. 

There is no way to connect planarly the open string to the determinant building the giant (as long as the impurities do not reach the boundary 
of the chain, which we are not considering here). As we see from the graph below, the leading contribution is given by $N^{J-1}N$. We have $J$ 
graphs of this type, coming from the different choices of the open bits to interact with the determinant. 
The total result for the diagram is:
\be
\frac{4\pi^2}{k^2}\left(\frac{N}{16\pi^2|x|^2}\right)^{N+J-1}(N-1)^2(N-2)!^3\,JN\,{\cal I}(|x|)=\frac{\lambda^2\,J}{N^4}\,\langle {\cal O}_n(x)
{\cal O}_n(0)\rangle_{tree}\, {\cal I}(|x|)\;,
\ee
which is subleading respect to (\ref{O-O}).

\begin{center}
\includegraphics[scale=0.5]{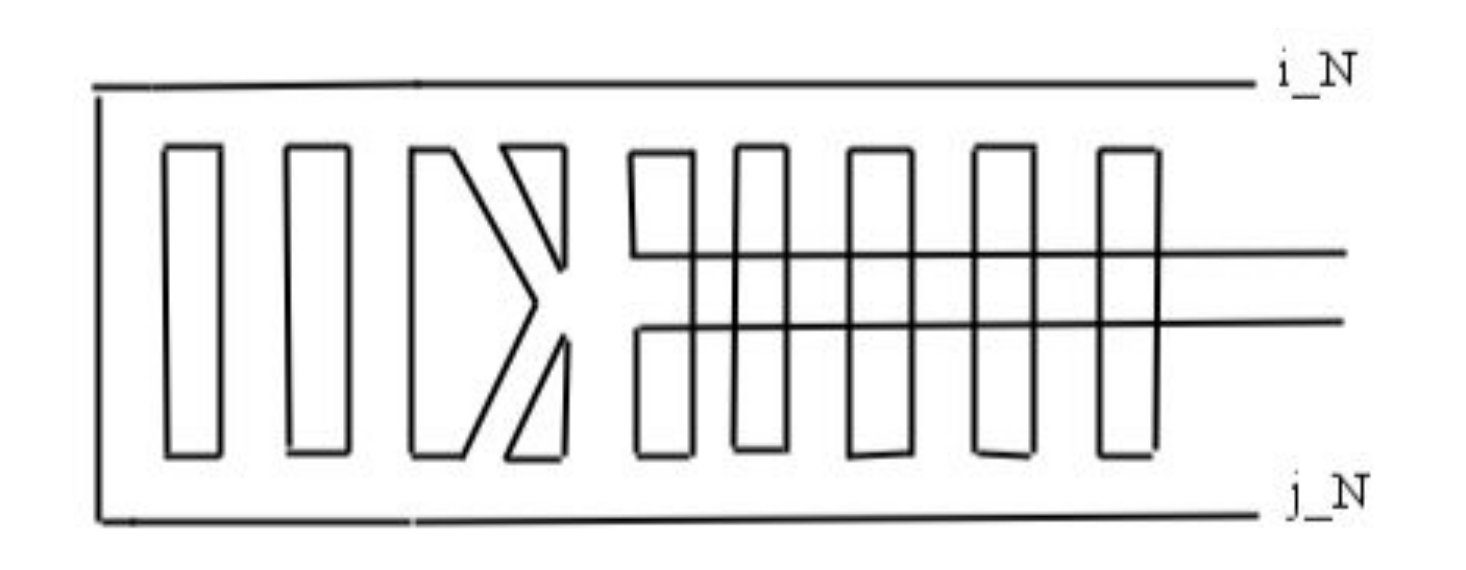}
\end{center}

\end{itemize}

\bibliography{OSinABJM3}{}
\bibliographystyle{utphys}

\end{document}